\documentclass[12pt,a4paper]{article}
\newcommand{\eqref}[1]{(\ref{#1})}
\begin{document}
\begin{center}
{\bf \large{An explicit realization of fractional statistics in one dimension}}\\
S. Sree Ranjani$^1$, P. K. Panigrahi$^2$ A. K. Kapoor$^3$ and A. Khare$^3$\\
$^{1,3}$School of Physics, University of Hyderabad, Hyderabad, 500 046, India.\\
$^2$Physical Research Laboratory, Ahmedabad, 380 009, India.\\
$^3$Institute of Physics, Bhubaneswar, 751005, India.\\
\end{center}

\begin{center}
{\bf Abstract}
\end{center}
 An explicit realization of anyons is provided, using the three-body
     Calogero model. The fact that in the coupling domain, $-1/4<g<0$, 
    the angular spectrum can have  a band structure, leads to the manifestation of the desired phase in the wave function, under the exchange of the paticles. Concurrently, the momentum  corresponding to the angular variable is
     quantized, exactly akin to the relative angular momentum
     quantization in two dimensional anyonic system.\\ 
\noindent     

\noindent
{\bf Introduction}

\noindent
      Fractional statistics (FS) has been widely explored in two spatial
      dimensions \cite{wil}, \cite{hal1}, \cite{khbook},
where it can arise due to the Abelian nature of the rotation group.
Furthermore, braid group replaces the permutation group in the plane
leading to the possibility of anyons \cite{wu}-\cite{sato}. The 
Chern-Simons (CS) theories in $2+1$ dimensions describe anyonic excitations \cite{polya}, \cite{pkp}, where an Abelian gauge field with CS action implements statistical interaction \cite{cho}-\cite{poly}. The corresponding 
quantum mechanical problem led to the statistical $1/r^2$ interaction in the
plane, which has been extensively studied \cite{polykro}. The possibility of self-adjoint
extensions and singular but normalizable wave functions, have been
explicated in models with statistical interaction \cite{asim}.
 In contrast to this, in one dimension, statistics and interaction get
non-trivially intertwined. However, unlike two spatial dimensions, here
solvable models exist, which show the effect of statistical interaction
explicitly. The Calogero-Sutherland model with its $1/r^2$ mutual
interaction between particles \cite{ha}, has shown the realization  of
Haldane statistics in one dimension in the coupling range $g>0$ \cite{gun} - \cite{isk}.  Here, the energy spectrum is linear with $E_{n_l}=\sum_l n_l +
\frac{N}{2}+\frac{N(N-1)}{2}(\alpha+\delta)$, where $n_l=0,1,2..$, $\alpha=\sqrt{(1+2g)}/2$ and $\delta$ is $0$ and $1$ for bosons and fermions respectively.  
Subsequently, an  explicit map has been
given relating two particle fermionic wave functions to the corresponding bosonic
ones \cite{mur3}. Later on, this has been generalized to N particles in an
oscillator potential \cite {basu}, as well as for free particle scenarios. 
In these studies, the wave functions show fermionic or bosonic
behavior by construction.

Here, we explicitly solve the three-body Calogero problem with
anyonic boundary conditions and establish FS through the phase acquired by the
wave functions. 
In the Jacobi co-ordinates, the three body Hamiltonian reduces
to a separable equation describing the dynamics of a particle on a plane, in a
non-central potential for which exact solutions exist. The solutions of the
angular part are central to this analysis, as the interchange of the particles
and their interaction are both reflected in the angular variable. By mapping
the potential in the angular equation to the Scarf potential \cite{scarf},
which can display band spectrum for a certain range of the potential parameter,
 anyonic boundary condition can be implemented on the wave
function. When the particle
positions are exchanged, this boundary condition allows the wave function to
pick up a phase, which is fractional  for solutions inside the bands and
integral for the band-edge solutions. We discuss the solutions for the
radial part briefly and present the complete solutions of the three body Calogero
model, in the parameter range $-1/4<g<0$. Thus this model
presents a unique situation, where the FS is established via the phase
acquired by the wave function. In addition to this, we also show that in the context of anyons, the quantized momentum associated with the band spectrum,  takes
the form similar to the quantization of angular momentum of anyons in two
dimensions \cite{wil} and the spacing of linear momentum in one dimensions discussed in
\cite{martin}.

   In the next section, we analyze the three-body Calogero model and map its
   angular eigenvalue problem to the Scarf potential. The solutions for the
   angular and radial equations are discussed and the complete solutions for
   the Calogero model are presented. In section III, we study the effect of
   the particle interchange on the solutions. In section IV, we discuss
   the quantization of the momentum in the angular variable. Here, we briefly discuss other three-body problems, where anyonic excitations can exist, followed by the conclusions in the last section. \\

\noindent
{\bf II Band structure problem in the Calogero model}\\
The Hamiltonian for the Calogero model \cite{cal} is $(\hbar = 2m = 1)$, 
\begin{equation} 
H=\sum_{i=1}^{3} -\frac{\partial^2}{\partial x^2_i} +\frac{\omega^2}{12}\sum_{i<j}(x_i -x_j)^2 +g\sum_{i<j}(x_i -x_j)^{-2},   \label{an1}
\end{equation}
where $g \geq -1/4$. Using the Jacobi co-ordinates,
\begin{equation}
X=\frac{(x_1+x_2+x_3)}{3}\,,\,x=\frac{(x_1-x_2)}{\sqrt{2}}\,,\, y=\frac{(x_1+x_2-2x_3)}{\sqrt{6}},  \label{an2}
\end{equation}
we factor out the center of mass of the three particles and obtain a system with two degrees of freedom. Next we map it on to the $(r, \phi)$ coordinates using
\begin{equation}
 x=r\sin\,\phi\,\,, \,\, y=r\cos\,\phi\,\,,\,\, r^2=\frac{1}{3} [(x_1-x_2)^2+(x_2-x_3)^2+(x_3-x_1)^2]. \label{an3}
\end{equation}
 The range of these variables is $0\leq r<\infty$ and $0\leq\phi<2\pi$ and
 the ordering of the particles is reflected in the variable $\phi$ \cite{cal}, for example
 the ordering $x_1>x_2>x_3$ corresponds to $0<\phi<\pi/3$. Thus
 \eqref{an1} reduces to,
\begin{equation}
\left (-\frac{\partial^2} {\partial r^2}-\frac{1}{r}\frac{\partial} {\partial r}
  -\frac{1}{r^2}\frac{\partial^2}{\partial \phi^2}\right) \Psi(r,\phi) =
\left(E - \tilde{U}(r) - \frac{U(\phi)}{r^2}\right) \Psi(r,\phi), \label{an4}
\end{equation}
where $ \tilde{U}(r) = \frac{1}{4}\omega r^2$ and
$U(\phi)=\frac{9g}{(\sin\, 3\, \phi)^2}$ \cite{kha}.  
Using 
\begin{equation}
\Psi(r,\phi)=\frac{1}{\sqrt{r}}u(r)K(\phi),  \label{anI}
\end{equation} 
we separate \eqref{an4} into the  radial and angular equations. The
 solutions of the radial equation,  
\begin{equation}
\left(-\frac{d^2}{dr^2}+ \tilde{U}(r)+\frac{(m^2-1/4)}{r^2} \right)u(r) =Eu(r),   \label{an5}
\end{equation}
are
\begin{equation}
u(r)= r^{m+\frac{1}{2}} \exp\left(-\frac{\omega r^2}{4}\right)\,
L^m_l\left(\frac{1}{2}\omega r^2\right)   \label{an14}
\end{equation}
where $L^m_l(r) $ are the Laguerre polynomials. The eigenvalues are
\begin{equation}
E=(2l+m+1)\omega,    \label{an15}
\end{equation}
where $m^2$ is the eigenvalue of the equation in $\phi$ given below
\begin{equation}
\left(-\frac{d^2}{d\phi^2} +\frac{9g}{(sin \,3\,\phi)^2} \right)K(\phi) = m^2 K(\phi).   \label{an6}
\end{equation}
In the literature, the solutions for the Calogero model have been largely
confined to $g>0$ regime \cite{cal} -\cite{guru}. Our 
interest lies in the solutions of \eqref{an1}, when $-1/4<g<0$ \cite{anjan}, 
because for $g$ lying in this range, \eqref{an6} exhibits band structure
resulting in non-normalizable solutions \cite{scarf}, \cite{sree}.   

This is expected because  for values of $g$ lying in this range,
$\frac{9g}{(sin \,3\,\phi)^2}$ is similar to the potential in a crystal
lattice and hence, the band structure. In terms of the $\phi$ variable, we can
explain this by applying the circular boundary condition, $K(0) =
K(2\pi)$.

    As pointed out earlier, the solutions of \eqref{an6} will
    play an important role in determining the type of statistics obeyed by
    the particles.
    Using the mapping $3\phi= \frac{\pi x}{a}$,  we can map
    \eqref{an6} to  
\begin{equation}
\left(-\frac{\pi^2}{a^2}\frac{d^2}{dx^2}-\frac{(\frac{1}{4} -s^2)}{(sin \frac{\pi x}{a})^2}
\right)K(x) = \lambda^2 K(x), \label{an7}  
\end{equation}
where $g = (\frac{1}{4} -s^2)$, $9 m^2 =\lambda^2$ and $a$ is the
lattice period. Owing to the constraint on $g$, we see that $s$ should lie in
the range $0<|s|<\frac{1}{2}$. This band structure problem has been discussed
in detail by Scarf in \cite{scarf}. 

    The most general solution for \eqref{an7}, written in the self-matching
form \cite{scarf},\cite{james}, is  
\begin{equation}
K(x) = \frac{r^N}{1+\sqrt{\rho}}\left[\frac{v(x-Na)}{v_0}
  +\sqrt{\rho}\frac{u(x-Na)}{u_0}\right]   \label{an8}
\end{equation}
for $(N-\frac{1}{2})a <x\leq (N+\frac{1}{2})a$, with $N=\frac{1}{2},
\frac{3}{2}....$. Here $v(x)$ and $u(x)$ are the linearly independent solutions
of \eqref{an7}, whose explicit form, for $0<x<a/2$, is as follows 
\begin{eqnarray}
u(x)&=&\sin^2\left(\frac{\pi x}{a}\right)^{(\frac{1}{4}+ \frac{|s|}{2})} \,
_2F_1\left[\frac{1}{4}+\frac{|s|}{2}+\frac{\lambda}{2},
  \frac{1}{4}+\frac{|s|}{2}-\frac{\lambda}{2};1+|s|; \sin^2\left(\frac{\pi
      x}{a}\right)\right], \nonumber\\
v(x)&=&\sin^2\left(\frac{\pi x}{a}\right)^{(\frac{1}{4}- \frac{|s|}{2})} \,
_2F_1\left[\frac{1}{4}-\frac{|s|}{2}+\frac{\lambda}{2},
  \frac{1}{4}-\frac{|s|}{2}-\frac{\lambda}{2};1-|s|; \sin^2\left(\frac{\pi
      x}{a}\right)\right] \nonumber \\ \label{an9}  
\end{eqnarray}
and 
\begin{equation}
r=(1+\sqrt{\rho})/(1-\sqrt{\rho})\,\,,
\rho=i\tan\left(\frac{ka}{2}\right). \label{anl9}
\end{equation}
We point out here that for $N=1/2$, the range of $x$
corresponds to the particle ordering $x_1>x_2>x_3$. The eigenvalue $m^2$ corresponding to
\eqref{an8} is 
\begin{equation}
m^2=\frac{1}{9 \pi^2}[\cos^{-1}(\sin\pi |s| \cos ka)]^2,  \label{an10}
\end{equation}
where the $n^{th}$ principal value of the inverse cosine function is to be
taken. Here, $k$ is the reduced wave number. For the $n^{th}$ band, it
takes values  $0$ and $\pi/a$ for the lower and upper band-edges respectively
and values in between them inside the band. Substituting \eqref{anl9} in \eqref{an8}, we obtain $K(x)$ in terms of $k$ as
\begin{equation} 
K_k(x))=\exp\left(iNka-\frac{ika}{2}\right)
\left[cos\left(\frac{ka}{2}\right)\frac{v(x- Na)}{v_0}   
  +i\sin\left(\frac{ka}{2}\right)\frac{u(x-Na)}{u_0}\right],    \label{ana8}
\end{equation}
where the subscript $k$ is used to show the dependence of the solution on the
wave vector. It can be seen easily from the above equation that $K(\phi)$
satisfies the Bloch condition:  
\begin{equation}
K_k(x+a) = \exp(ika)K_k(x),   \label{an10a}
\end{equation}
using which, we can obtain the solutions for the entire lattice.
   Substituting  $k =0$ and $ k=\pi/a$ in \eqref{an10} we obtain the band-edge
eigenvalues for the lower and upper band-edges as 
\begin{equation} 
m^2_{0}=\frac{1}{9\pi^2}\left(n-|s|+\frac{1}{2}\right)^2\,,\,m^2_{\pi/a}=\frac{1}{9\pi^2}\left(n+|s|+\frac{1}{2}\right)^2.
\label{ane11} 
\end{equation}
Similarly for these values of $k$, $K(x)$ in \eqref{ana8} becomes pure 
$v(x)$ and pure $u(x)$ for the lower and upper band-edges respectively and
the hypergeometric functions in these solutions reduce to the Jacobi
polynomials \cite{sree}. Thus for the $n^{th}$ band, the lower band-edge eigenfunction
is
\begin{equation}
K_{0}(x) =\exp\left(iNka-\frac{ika}{2}\right)\left[\sin^2\left(\frac{\pi
    x}{a}\right)^{\frac{n}{2}-\frac{|s|}{2}+\frac{1}{4}}\right]P_n^{\nu_1,\nu_1}\left(-i\cot\left(\frac{\pi x}{a}\right)\right),    \label{and12}   
\end{equation}
where $\nu_1=-n+ |s|-\frac{1}{2}$
and the upper band-edge solution is 
\begin{equation}
K_{\frac{\pi}{a}}(x)
=\exp\left(iNka-\frac{ika}{2}\right)\left[\sin^2\left(\frac{\pi 
    x}{a}\right)^{\frac{n}{2}+\frac{|s|}{2}+\frac{1}{4}}\right]
  P_n^{\nu_1,\nu_1}\left(-i\cot\left(\frac{\pi x}{a}\right)\right),   \label{and13} 
\end{equation}
where $\nu_1=-n-|s|-\frac{1}{2}$.
Applying the Bloch condition to the band-edge solutions \eqref{and12} and \eqref{and13}, we
see that $K_0(x)$ does not acquire a phase and $K_{\frac{\pi}{a}}(x)$ changes sign when one
goes from one cell to the other. We would like to point out here that the
above solutions \eqref{an8},\eqref{ana8}, \eqref{and12} and \eqref{and13} are
defined in only one lattice cell. 
In the $\phi$ variable this corresponds to the range
$0\leq\phi<\pi/3$. Incrementing $\phi$ by $\phi \rightarrow \phi+p \pi/3$,
where $p=1,2,3,4,5$  defines the solutions for the entire range of $\phi$
\cite{cal} and this in the $x$ variable, amounts to translating from one cell
to the next cell in the lattice. Thus, by continuously incrementing $\phi$ by
$\pi/3$, we can cover the whole real line in the $x$ variable. 

 Having described the solutions for
both the radial and the angular eigenvalue equations, we write the complete
solution  of \eqref{an4} as
\begin{equation}
\Psi(r,\phi) = r^{m} \exp\left(-\frac{\omega r^2}{4}\right)\,
L^m_l\left(\frac{1}{2}\omega r^2\right) K_k(\phi),   \label{anr13}
\end{equation}
where we substitute the suitable solutions $K_k(\phi)$, with the corresponding
$m$  values, depending on the value of $k$ being considered. 
For $k$ lying inside the band, the complete 
  solution for \eqref{an4} is 
\begin{eqnarray}
\Psi(r,\phi) =  r^{|m|} \exp\left(-\frac{\omega r^2}{4}\right)\,
L^{|m|}_l\left(\frac{1}{2}\omega r^2\right)K_k(3\phi), \label{an16}
\end{eqnarray}
where the $m$ values are obtained from \eqref{an10}.

   Similarly for $k=0$ and $k=\pi/a$, the complete solution $\Psi(r,\phi)$ is
\begin{eqnarray}
\Psi(r,\phi) = r^{|m_0|} \exp\left(-\frac{\omega r^2}{4}\right)\,
L^{|m_0|}_l\left(\frac{1}{2}\omega r^2\right)K_{0}(3\phi) \label{an17}
\end{eqnarray}
and 
\begin{eqnarray}
\Psi(r,\phi) = r^{|m_{\pi/a}|} \exp\left(-\frac{\omega r^2}{4}\right)\,
L^{|m_{\pi/a}|}_l\left(\frac{1}{2}\omega r^2\right)K_{\frac{\pi}{a}}(3\phi), \label{an18}
\end{eqnarray}
where $m$ values are obtained from \eqref{ane11}. Thus we can see that for
$-1/4<g<0$,  the Calogero model has different solutions, depending on the wave vector, owing to the lattice like behaviour of the potential in the $\phi$ variable. For completeness, we point out here that for $g>0$, the complete solutions of the Calogero model are a product of the Gegenbauer polynomials comming from the angular part and the Laguerre polynomials  comming from the radial part \cite{cal}, \cite{kha}.  

\noindent
{\bf III Behaviour of the wave function under particle exchange}\\
   As mentioned earlier, the interchange of particles is reflected in the $\phi$ variable.  
    The range $0<\phi<\pi/3$, corresponding to the particle ordering
    $x_1>x_2>x_3$, leads to $0<x<a$ in the $x$ variable,  which corresponds
    to the first cell  of  the lattice. The change in particle  
    ordering, brought about by changing $ \phi\rightarrow
    \phi+p\pi/3$ with $p=1,2,3,4,5$, translates to
    moving from  one cell to the next. This leads to the wave function
    picking up a  phase as can be seen easily from \eqref{ana8}. The type
    of phase acquired, namely integral or fractional, will decide the
    statistics obeyed by the  system. 
      
   For explicitness, we change the ordering from $x_1>x_2>x_3$ to
   $x_2>x_1>x_3$   which is obtained by incrementing $\phi \rightarrow
   \phi+2\pi/3$. The range $\pi/3<\phi<2\pi/3$ corresponds to $a<x<2a$ which
   is the second  cell. Thus an interchange of particles requires an increment
   in $\phi$,
   which in $x$ variable corresponds to a translation to the second cell. The
   solutions in the first and second cell are obtained by putting $N=1/2$ and
   $N=3/2$  in \eqref{ana8} as 
   \begin{equation}
   K_{1,\,\,k}(x) = cos\left(\frac{ka}{2}\right)\left[\frac{v(x-
       \frac{a}{2})}{v_0} \right]
   +i\sin\left(\frac{ka}{2}\right)\left[\frac{u(x-\frac{a}{2})}{u_0}\right],
   \label{an19} 
  \end{equation}
with 
    \begin{eqnarray}
   K_{2,\,\,k}(x) &=& \exp(ika) cos\left(\frac{ka}{2}\right)\left[\frac{v(x-
       \frac{3a}{2})}{v_0}\right]
   +i\sin\left(\frac{ka}{2}\right)\left[\frac{u(x-\frac{3a}{2})}{u_0}\right],   \nonumber\\
    &=& \exp(ika)K_{1,\,\,k} (x).   \label{an20}
  \end{eqnarray}
 Here the additional subscripts, $1$ and $2$, in the left hand side of \eqref{an19} and \eqref{an20} denote the number of the cell. Thus when we translate from the first cell to the second cell of the lattice, we see that the wave function  picks up a phase 
     \begin{equation}
\Theta=ka,  \label{an21}
\end{equation}
 which we define as the statistics parameter. $\Theta$ takes values lying in the range
 $0\leq\Theta\leq\pi$ for $0\leq k \leq \pi/a$.  It
 is easy to see that every time the ordering of the particles is changed, the
 wave function picks up a phase which is fractional when $k$ lies in the
 range $0<k<\pi/a$. Thus for $k$ values  lying inside the band, the
 particles obey anyonic statistics. For $k=0$, which corresponds to the lower
 band-edge solution \eqref{and12}, $\Theta=0$, which implies the translation from first cell to
 the second cell does not affect the wave function {\it i.e.,} the lower
 band-edge solutions are symmetric implying a compliance  to bosonic
 statistics. On the contrary, for
 $k=\pi/a$, which corresponds to the upper band-edge solution \eqref{and13},
 $\Theta = \pi$, showing that the interchange leads to the wave
 function being antisymmetric, which implies that fermionic statistics are
 obeyed. Moreover, substituting $ka=\Theta$ in \eqref{an10}, we can see that
 the angular eigenvalues are  functions of the statistics parameter and for a
 given band they continuously interpolate between the bosonic and fermionic
 eigen energies. This interpolation between symmetrization and
 antisymmetrization  has been discussed by Wu in the context of three  particles in an harmonic well in two dimensions \cite{wu2} and by Yang {\it et. al.,} in one dimension for a Bose gas with two-body delta-function interactions \cite{yang}. 
 
    Thus, we can see that, the band structure in the eigenvalue spectrum of the
 differential equation in $\phi$ leads to rich physics, where the particles
 obey three different statistics depending on the value of $k$. It is  interesting to note that in \cite{mur1}, the CSM  $(g>0)$, with N-body interaction,  has been shown to obey FS as defined by Haldane.  In this case the energy spectrum is linear and the scale invariant energy shift is seen as the basic reason  for the occurence of the Haldane statistics. This result holds for the three-body Calogero model also as its energy spectrum is linear \cite{cal}, in the $g>0$ regime. When we compare this to the present case, we see that for $-1/4<g<0$, the spectrum given by \eqref{an15} in nonlinear owing to the fact that $m$, given by \eqref{an10}, is a non-integer, where fractional statistics manifest.  The fractional phase is picked up by the wave function under particle exchange. We expect this result to hold when our analysis is extended to the  general CSM with $-1/4<g<0$. 
 

\noindent 
{\bf IV Quantization of the momentum in the $\phi$ variable} 

   Exploiting the well developed theory of energy bands in solid state
   physics, we show that the momentum in the $\phi$ variable, corresponding to
   the anyonic statistics, is quantized. This is similar to the relative
   angular momentum quantization, proposed by Wilczek, for anyons in
   two dimensions \cite{wil}.   
   
   The effective crystal momentum associated with the particle in a lattice is
   defined as
   \begin{equation}
   p(k)=\frac{\hbar}{ia}\ln[r(k)].  \label{an22}
   \end{equation}
   Substituting $\rho=i\tan\left(\frac{ka}{2}\right)$ for $r$ in \eqref{anl9},
   gives $r=\exp(ika)$, which in turn gives 
   \begin{eqnarray}
   p&=&\frac{\hbar}{ia}\ln (\exp[ika+2n^{\prime}\pi i])  \nonumber \\
    &=&\frac{h}{a}[\frac{ka}{2\pi} +n^{\prime}].\label{an23}
   \end{eqnarray} 
   Here $n^{\prime}$ is a non-negative integer. Substituting $ka=\Theta$ from
   \eqref{an21}, we get 
   \begin{equation}
   p=\frac{h}{a}[\frac{\Theta}{2\pi} +n^{\prime}],\label{an22}
   \end{equation}
which is the  quantization condition for $p$ in terms of the statistics parameter.
It is interesting to note that we can extend this analysis and establish fractional statistics in other exactly solvable three-body problems discussed by Wolfes \cite{wolfes}, Khare and Bhaduri \cite{KH3body}, where the angular eigenvalue equation will exhibit band spectrum for a small range of potential parameters. Similar analysis of these problems shows that inside the band, the particles obey anyonic statistics and at the band edges, the statistics obeyed are bosonic and fermionic. We briefly discuss the Wolfes potential here.

\noindent
{\bf  Wolfes potential}

The eigenvalue equation for the three-body linear problem discussed by Wolfes \cite{wolfes} is 
\begin{eqnarray}
\left(-\sum_{i=1}^3 \frac{\partial^2}{\partial x_i^2} +\sum_{i,j=1}^3[\frac{1}{6}\omega^2(x_i -x_j)^2 + 2g(x_i-x_j)^{-2} ]\right)\Psi_{i \neq j\neq k}(x_1,x_2,x_3)  \nonumber  \\ 
+\left(\sum_{i,j,k=1}^3 6f[(x_i-x_j)+(x_j-x_k)]^{-3} - E\right)  \Psi_{i \neq   j\neq k}(x_1,x_2,x_2)   = 0,   \label{an23}
\end{eqnarray}
where $g>-1/4$ and $f>-1/4$. In the polar coordinates, this model can be separated to obtain 
\begin{equation}
\left(-\frac{\partial^2}{\partial x^2} -\frac{1}{r}\frac{\partial}{\partial r}+\omega^2r^2 +\frac{\lambda^2}{r^2} -E\right) \chi(r) =0    \label{an24}
\end{equation}
and 
\begin{equation}
\left(-\frac{\partial^2}{\partial \phi^2} + \frac{9g}{\sin^2 3\phi} +\frac{9f}{\cos^2 3\phi}-\lambda^2\right) K(\phi) = 0.    \label{an25}
\end{equation}
Similar to the Calogero model studied earlier, the angular equation exhibits band spectrum for $-1/4<g<0$ and $-1/4<f<0$. Using the mapping $3\phi =\frac{\pi x}{a}$ and defining $t=\sin^2(\frac{\pi x}{a})$, $g=(s^2 -\frac{1}{4})$ and $f=(w^2 -\frac{1}{4})$, we can map the angular equation \eqref{an25} to an equation similar to \eqref{an7} 
\begin{equation}
\left(-\frac{a^2}{\pi^2}\frac{d^2}{d\phi^2} +\frac{9g}{\sin^2(\frac{\pi x}{a})}+\frac{9f}{\cos^2(\frac{\pi x}{a})} -\lambda^2\right)K(\phi) =0.   \label{an26}
\end{equation}
By substituting  $\phi(z) = t^{\alpha}(1-t)^bF(t)$ in \eqref{an26}, it can be reduced to a hypergeometric equation,
whose linearly independent solutions in terms of $x$ are
\begin{eqnarray}
&u(x)&= \left[\sin^2\left(\frac{\pi x}{a}\right)\right]^{\frac{1}{4}+\frac{|s|}{2}}\left[\cos^2\left(\frac{\pi x}{a}\right)\right]^{\frac{1}{4}+\frac{|w|}{2}} \nonumber \\ &&_2F_1\left[\frac{1}{2} +\frac{|s|}{2} +\frac{|w|}{2}-\frac{\lambda}{6}, \frac{1}{2} +\frac{|s|}{2} +\frac{|w|}{2}+\frac{\lambda}{6}; 1+|s|; \sin^2\left(\frac{\pi x}{a}\right)\right] \nonumber\\ \label{an27}
\end{eqnarray}
and 
\begin{eqnarray}
&v(x)&= \left[\sin^2\left(\frac{\pi x}{a}\right)\right]^{\frac{1}{4}+\frac{|s|}{2}}\left[\cos^2\left(\frac{\pi x}{a}\right)\right]^{\frac{1}{4}+\frac{|w|}{2}} \nonumber \\ 
&&_2F_1\left[\frac{1}{2} -\frac{|s|}{2} -\frac{|w|}{2}-\frac{\lambda}{6}, \frac{1}{2} -\frac{|s|}{2} -\frac{|w|}{2}+\frac{\lambda}{6}; 1-|s|; \sin^2\left(\frac{\pi x}{a}\right)\right]\nonumber\\  \label{an28}
\end{eqnarray} 

Proceeding in the same way as in the Calogero case, the above two solutions can be used to write the most general solution in the self-matching form, for $(N-1/2)<x<(N+1/2)$, given in \eqref{an8}. As expected, at the band-edges, the self-matching solution becomes pure $v(x)$ or $u(x)$ with the hypergeometric functions reducing to the Jacobi polynomials. The solution $K(x)$, with $v(x)$ or $u(x)$ substituted from \eqref{an27} and \eqref{an28}, picks up a phase $\Theta$, when we travel from one cell to the other. As discussed earlier, in the $\phi$ variable, this translates to exchanging the particle positions. Not suprisingly, $\Theta$ is fractional for $0<k<\pi/a$ and $\pm 1$ for $k=0$ and $k=\pi/a$ respectively.  Thus, it is sufficient to conclude that the three-body problem described by the Wolfes potential will also show fractional statistics in the angular equation. From above, it is clear that this analysis can also be extended to similar three-body interactions discussed by Khare and Bhaduri in \cite{KH3body}, as the angular equation in all these problems also displays band spectrum in some range of the potential parameters. Thus, the fractional statistics seems to be a common feature  for all the three-body interactions, whose angular potential exhibits lattice like behaviour for  some parameter range of the potential.

\noindent
{\bf VI Conclusion}

  In this paper, we have analyzed the solutions of the Calogero model for $-1/4<g<0$ and we
have shown that the band structure in the angular variable has
interesting consequences for the statistics obeyed by the
particles. The angular eigenvalues turn out to be functions of the statistics
parameter $\Theta$, which is the phase picked up by the wave function when two particles are exchanged. 
For $\Theta$ lying in the range $0<\Theta<\pi$, the statistics obeyed 
is anyonic. For $\Theta=0$ and $\Theta=\pi$,  the particles obey bosonic and
fermionic statistics respectively. We have also shown that in a given band, there
is a continuous interpolation from the bosonic state to the fermonic state.  In addition to this,
the quantization of momentum in the $\phi$ variable, in terms of the statistics parameter, is obtained. 

 We have also showed that this type of statistical behaviour is common for the whole class of three-body problems, whose angular equation  displays a band spectrum.  It will be interesting to extend this study to models with $N>3$ and finally to the $N$-body CSM in the parameter regime $-1/4<g<0$. The possible difficulty would be the nonavailability of exact solutions in this regime. Hence, we may have to resort to the use of numerical techniques and this will be discussed else where.
     Here, we would like to emphasize two points. First, the anyonic phase picked up by the wave function is due to the exchange of the particles.  Though the statistics here are anyon-like, these are not analogues of the braid statistics obeyed by anyons in two dimensions \cite{wil} and hence there are no braids associated with these statistics. Secondly, these statistics are entirely different from Haldane's fractional exclusion statistics which are based on the generalized  Pauli's exclusion principle \cite{hal1}. As mentioned earlier, such statistics is obeyed by the Calogero model in the $g>0$ regime and the main reason for this non-trivial statistics here, is the linearity of the energy spectrum.  
   
The intense research on Bose-Einstein condensates (BECS), has led to the development of experimental techniques to control and manipulate a few atoms in suitable traps. In light of these developments, it looks feasible for the above anyon model to be realized experimentally \cite{cirac}. It is interesting to note that there have been proposals to use anyons for quantum computation \cite{kit}.  Hence, the realization of  FS in one dimension, in the class of three body problems, may have useful applications in the area of quantum computation and in the engineering of entangled states of atoms.\\

\noindent
{\bf References}


\end{document}